\documentclass[twocolumn]{aastex63}

\usepackage{natbib}

\accepted{to ApJ, September 2021}

\begin{document}

\title{Probing Current Sheet Instabilities from Flare Ribbon Dynamics}

\author{Ryan J. French}
\affil{Mullard Space Science Laboratory, University College London, Dorking, RH5 6NT, UK}

\author{Sarah A. Matthews}
\affil{Mullard Space Science Laboratory, University College London, Dorking, RH5 6NT, UK}

\author{I. Jonathan Rae}
\affil{Department of Mathematics, Physics and Electrical Engineering, Northumbria University, Newcastle upon Tyne, UK}

\author{Andrew W. Smith}
\affil{Mullard Space Science Laboratory, University College London, Dorking, RH5 6NT, UK}


\begin{abstract}

The presence of current sheet instabilities, such as the tearing mode instability, are needed to account for the observed rate of energy release in solar flares. Insights into these current sheet dynamics can be revealed by the behaviour of flare ribbon substructure, as magnetic reconnection accelerates particles down newly reconnected field lines into the chromosphere to mark the flare footpoints. Behaviour  in  the  ribbons  can  therefore  be  used  to  probe processes  occurring  in  the  current  sheet. 

In this study, we use high-cadence (1.7 s) IRIS Slit Jaw Imager observations to probe for the growth and evolution of key spatial scales along the flare ribbons - resulting from dynamics across the current sheet of a small solar flare on December 6th 2016. Combining analysis of spatial scale growth with Si IV non-thermal velocities, we piece together a timeline of flare onset for this confined event, and provide evidence of the tearing-mode instability triggering a cascade and inverse cascade towards a power spectrum consistent with plasma turbulence.

\end{abstract}


\section{Introduction} 

In the standard eruptive flare model, magnetic reconnection originates in a thin current sheet; created by the inflow of oppositely orientated magnetic fields under a rising magnetic flux rope \citep{Carmichael,Sturrock,Hirayama,Kopp}. The nature of magnetic reconnection in the current sheet, and the processes leading to their formation, are not yet fully understood. In order to explain the breakdown of ideal magnetohydrodynamics (on small spatial scales) needed to enable fast magnetic reconnection in the current sheet (leading to solar flare onset), recent attention has turned to the role of the tearing-mode instability \citep{Biskamp1986}. In the tearing-mode instability, a classic Sweet-Parker current sheet reaches an unstable aspect ratio, collapsing and reconnecting at multiple points along the sheet to form magnetic islands, or plasmoids. These magnetic islands form at a preferred spatial scale, and continue to collapse further to produce islands at progressively smaller scales. Larger islands are also produced via the reconnection and coalescence at island boundaries \citep{Tenerani2020}. This critical aspect ratio and spatial scales are dependent on the local magnetic conditions, in particular the Lunduist number. The cascade to smaller island scales is predicted to produce plasma turbulence, with a power-law similar to those observed in in-situ plasmas \citep{Shaikh2010}. The exact relationship between the tearing-mode instability and turbulence within a reconnecting current sheet is still a prominent area of ongoing research.

Current sheets are notoriously difficult to observe directly, primarily due to the low density and small width of the region. Heated regions of plasma around current sheets (known as a plasma sheet) can occasionally be observed off-limb, but the rarity of their observation means that spectroscopic and multi-wavelength observations, needed to truly disentangle the energy release process (and  relationship with instabilities and plasma turbulence), are rare. The most successful spectroscopic plasma sheet observation set to date originated from the famous September 10th 2017 X-class flare \citep{Warren}. This event was observed by multiple instruments across the entire electromagnetic spectrum, facilitating insight into the nature of magnetic reconnection in the event, and providing initial evidence for the presence of turbulent or tearing-mode reconnection \citep{Cheng, French2019}, persisting for the flare's entire duration \citep{French2020}. Although observations of this September 10th 2017 event provided many new insights into the behaviour of current sheet dynamics, the uniqueness of such a dataset requires alternative methods to be employed in order to make further progress in understanding these processes. The field line connectivity between flare ribbons that are routinely observed on disk and the reconnecting current sheet, offers such a possibility.
On-disk observations of flares are far more common, as active regions can be observed for longer durations on-disk than they can off-limb.

During a solar flare, high energy particles are accelerated from the reconnection site down the flare loops to the chromosphere, depositing energy at the magnetic footpoints to form flare ribbons \citep{Cheng1983,Doschek1983}. Due to their magnetic connectivity, behaviour of flare ribbon substructure must in some way reflect processes in the flaring current sheet region \citep{Forbes2000}. Observational studies of flare ribbons have found spectroscopic signatures suggestive of waves and/or turbulence \citep{Brosius2015} - consistent with the presence of either the tearing mode or Kelvin-Helmholtz instability in the current sheet \citep{Brannon2015}. These signatures of energy release in the chromosphere are detected on as fine a scale as $\sim$ 150-300 km \citep{Graham2015,Jing2016}. 

The Earth’s magnetotail experiences similar cycles of energy storage and release \citep{Akasofu1964,McPherron1970}. The energy builds in the Earth’s magnetotail until it reaches a point of instability, at which point the energy is explosively released through reconnection \citep{Hones1984}, causing observable phenomena such as the rapid brightening of the aurora \citep{Akasofu1977} (analogous to flare ribbon brightening on the Sun), and the formation of auroral beads \citep{Henderson1994,Henderson2009}. Fourier analysis of auroral bead brightenings along the auroral arc has found that specific spatial scales grow exponentially \citep{Rae2010,Kalmoni2015}. Recently, high temporal and spatial resolution measurements of the aurora enabled the comparison of observation and theory to show that non-local effects, in the form of sheer Alfven waves, were responsible for plasma instabilities in near-Earth space \citep{Kalmoni2018}.

Although fundamentally different in their temperature, density and collisional timescales, there are distinct similarities between flare ribbons and terrestrial aurorae driven by sub-storm activity. In this study, we take inspiration from the magnetospheric work of \citet{Kalmoni2018}, applying the auroral beads methodology to flare ribbons of a small solar flare, to determine whether similar evidence exists for the presence of a plasma instability at flare onset.

\section{Observations} 

We present IRIS \citep{DePontieu} observations of a confined B-class flare in the core of AR 12615 on December 6th 2016. IRIS observed with a \textit{large sit-and-stare} Slit Jaw Imager (SJI) 1400 \AA\ window, and a 1.7 s cadence. With just the one slit-jaw channel, this is close to the fastest cadence IRIS can observe. The SJI observations were 2x2 binned, resulting in a pixel width of 0.3327 \arcsec\ or 238 km.
IRIS observed the flare continuously from pre-flare to decay.

\begin{figure*}
  \centering
  \includegraphics[width=16cm]{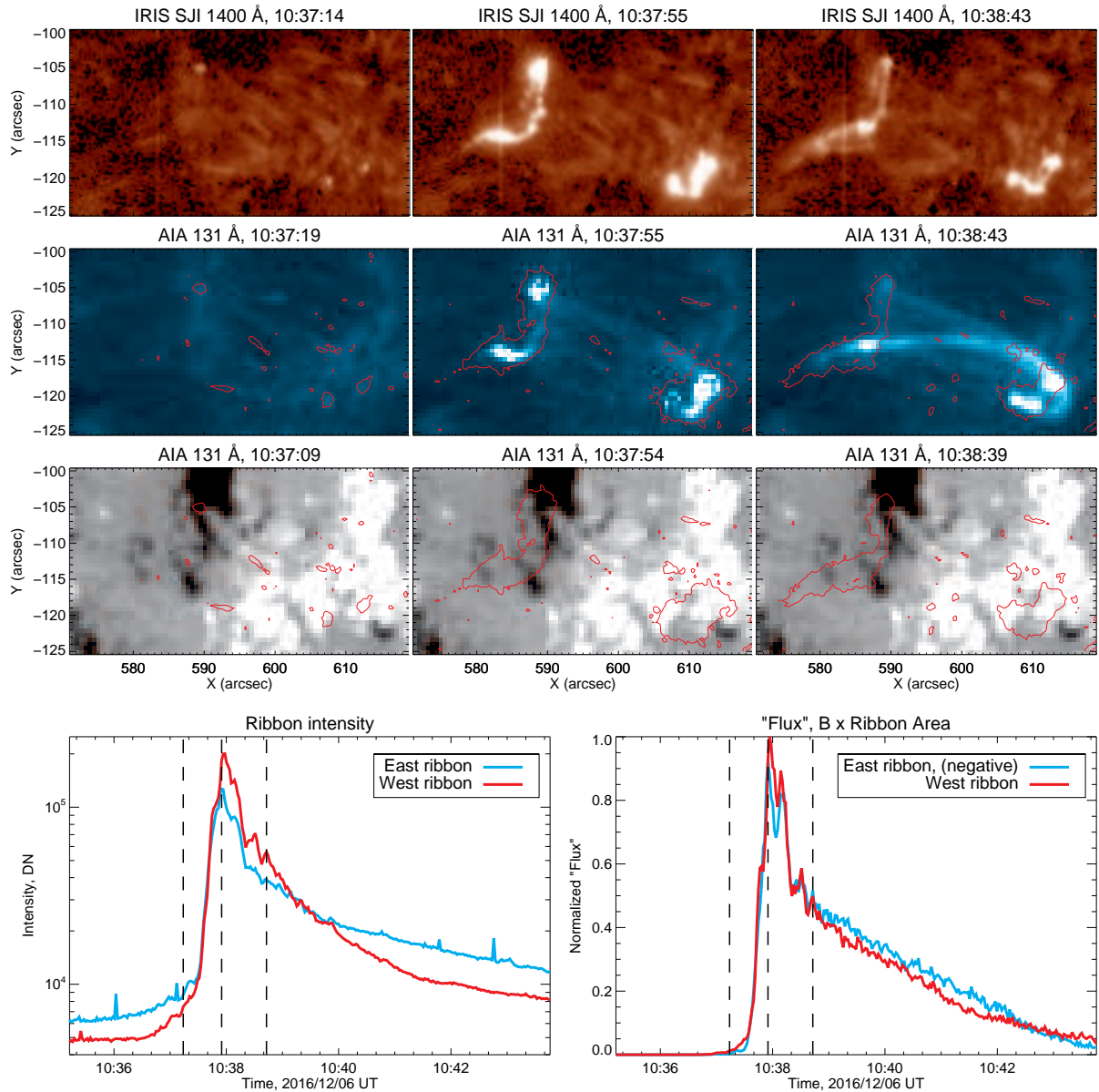}
  \caption{
Top: Evolution of IRIS SJI 1400 Å. 
Second row: Evolution of AIA 131 Å, with IRIS ribbon contours to distinguish emission at high/low altitudes. 
Third row: Evolution of HMI LoS B field, with IRIS ribbon contours.
Bottom Left: Ribbon light curve, of masked region in Fig \ref{fig:tracking}. Vertical dashed lines mark location of above panels.
Bottom Right: Ribbon area over intensity threshold, multiplied by HMI LoS field strength in each region. This is analogous to magnetic flux (BxA) evolution within the bright ribbon region. East ribbon flux is multiplied by -1. 
}
  \label{fig:ribbons}
\end{figure*}

Figure \ref{fig:ribbons} (top row) presents snapshots of the SJI 1400 \AA\ evolution, with the field of view cropped to center on the two flare ribbons. From hereon, the two ribbons will be denoted as the \textit{east} (left) and \textit{west} (right) ribbon. 
Examining the light curve of each region, we see the two ribbons brighten cotemporally, and although the east ribbon is significantly larger than the west, maximum intensities are comparable.
Similar maximum intensities imply a similar level of energy deposition in each ribbon.

The brightest ribbon regions are also observed in the 131 \AA\ channel of the Solar Dynamics Observatory Atmospheric Imaging Assembly \citep[AIA,][]{Lemen2012}. Following peak ribbon intensity, higher-altitude flare loops become visible in this channel as they fill with hotter ablated chromospheric material. The timing of these events is shown in Figure \ref{fig:ribbons}. Looking at the photospheric Line of Sight (LOS) magnetic field measurements (B) in this region from the Helioseismic and
Magnetic Imager \citep[HMI,][]{Schou2012}, we see the flare ribbons (marked by red contour) appear either side of the polarity inversion line, adjacent to the regions of highest field strength (third row of Figure \ref{fig:ribbons}). At around 600 \arcsec\ from center disk, projection effects have begun to affect the west of the HMI observations. Although the area traced by the ribbons change throughout their evolution, the photospheric field itself does not vary on the timescales studied. Examining the LoS magnetic flux traced by the evolving (SJI 1400 \AA) ribbons in the bottom-right panel of Figure \ref{fig:ribbons}, we note that the flux is equal in each ribbon throughout the flare. That is,

\begin{equation}
  B_{E}A_{E} = B_{W}A_{W}
  \label{eqn:flux1}
\end{equation}

Matching flux in each ribbon implies connection of the two via a consistent tube of flux, which does not interact significantly with any external field. We can therefore imagine this flux tube passing from one ribbon to the other, containing the reconnection regions within it.  The flare is likely a result of quadrapolar reconnection \citep{Melrose1997,Pikelner1977}, with, due to the small size of the event, footpoints of equal polarity merging together to form one ribbon feature either side of the polarity inversion line. We therefore expect the reconnecting current sheet to be somewhere between the flare loops - perhaps even those seen in AIA 131 \AA. 
The Active Region is complex however, and other flaring configurations are possible. One other possibility considered was the circular ribbon flare \citep{Wang2012}, which would feature null point reconnection instead of the more-traditional laminar current sheet. In this scenario, null point reconnection still contains layers of current and thus would still be subject to instabilities \citep{Pontin2011}.

Spectral data of the flare ribbons were collected by the IRIS sit-and-stare slit, centered across the larger (east) ribbon. The position of this vertical slit can be seen as a slight enhancement in SJI image intensity.
\citet{Jeffrey2018} examine the spectral evolution of this flare ribbon, finding a steep rise in Si IV 1402.77 \AA\ non-thermal velocity (interpreted as a signature of plasma turbulence) preceding the rise in Si IV intensity (marking plasma heating flare onset).

In our study, we examine the spatial and temporal evolution of the two ribbons in SJI 1400 \AA\ observations, investigating the power spectra in the spatial domain along the ribbon to provide insight into processes across the current sheet, perpendicular to the magnetic field. We compare the power spectra at multiple spatial scales with parameters predicted by theory, and the timings of the Si IV non-thermal velocities introduced by \citet{Jeffrey2018}. When studying the ribbon spatial intensity evolution, we are not restricted by the slit location and can therefore analyze both ribbons independently. 
Current sheet processes may also affect structure across the ribbon, in addition to the variation along the ribbon we probe in this study. Analyzing these components together, along with the time-varying component, would provide the full dispersion relation, but not possible with the resolution of this data set. We therefore focus exclusively on searching for spatial scales along the flare ribbon structures.

\section{Ribbon Power Spectrum}
\subsection{Ribbon Tracking and Intensity Processing}

\begin{figure}
  \centering
  \includegraphics[width=8.5cm]{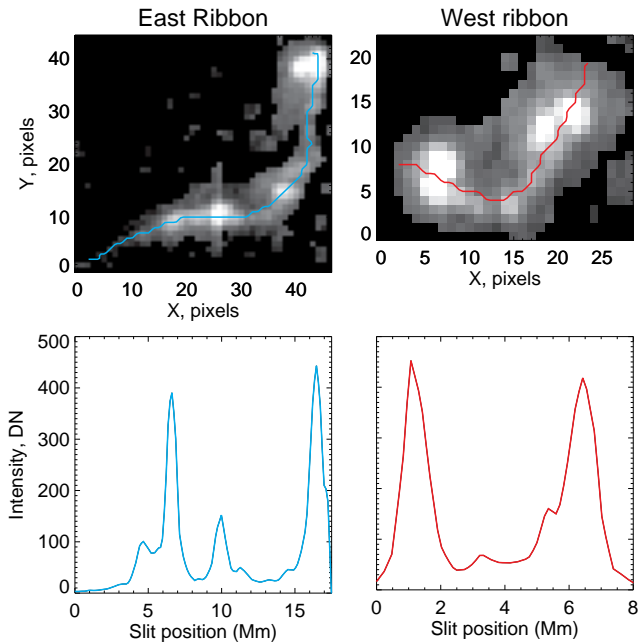}
  \caption{
Top: Snapshots (at 10:37:37 UT) of the ribbon tracking centroid position, along the masked data of the each ribbon. 
Second Row: Intensity cross-sections along the centroids in the above panel.
  }
  \label{fig:tracking}
\end{figure}

In order to investigate intensity variation along the flare ribbons, we developed a method to track a centroid along each flare ribbon as they evolved in space and time. This method, and all subsequent analysis, is repeated separately for both the east and west ribbon.

To begin, we produced a mask to remove all pixels never crossing an intensity threshold (that never form part of the ribbon structure). Next, we take cross-sections perpendicular to the main ribbon structure, and locate the position of the brightest pixel at each of these cross-sections. We then take a moving average of the pixel locations from this initial estimate, producing a smooth, continuous centroid slit along the centroid of each ribbon. These centroids change as the ribbons evolve, and a snapshot for each ribbon is shown in the top row of Figure \ref{fig:tracking}. Our main tracking limitation is the spatial resolution, as each centroid position must still have an integer pixel position. The tracking method works effectively however, capturing the change in curvature of the ribbons as they evolve.

\begin{figure*}
  \centering
  \includegraphics[width=16cm]{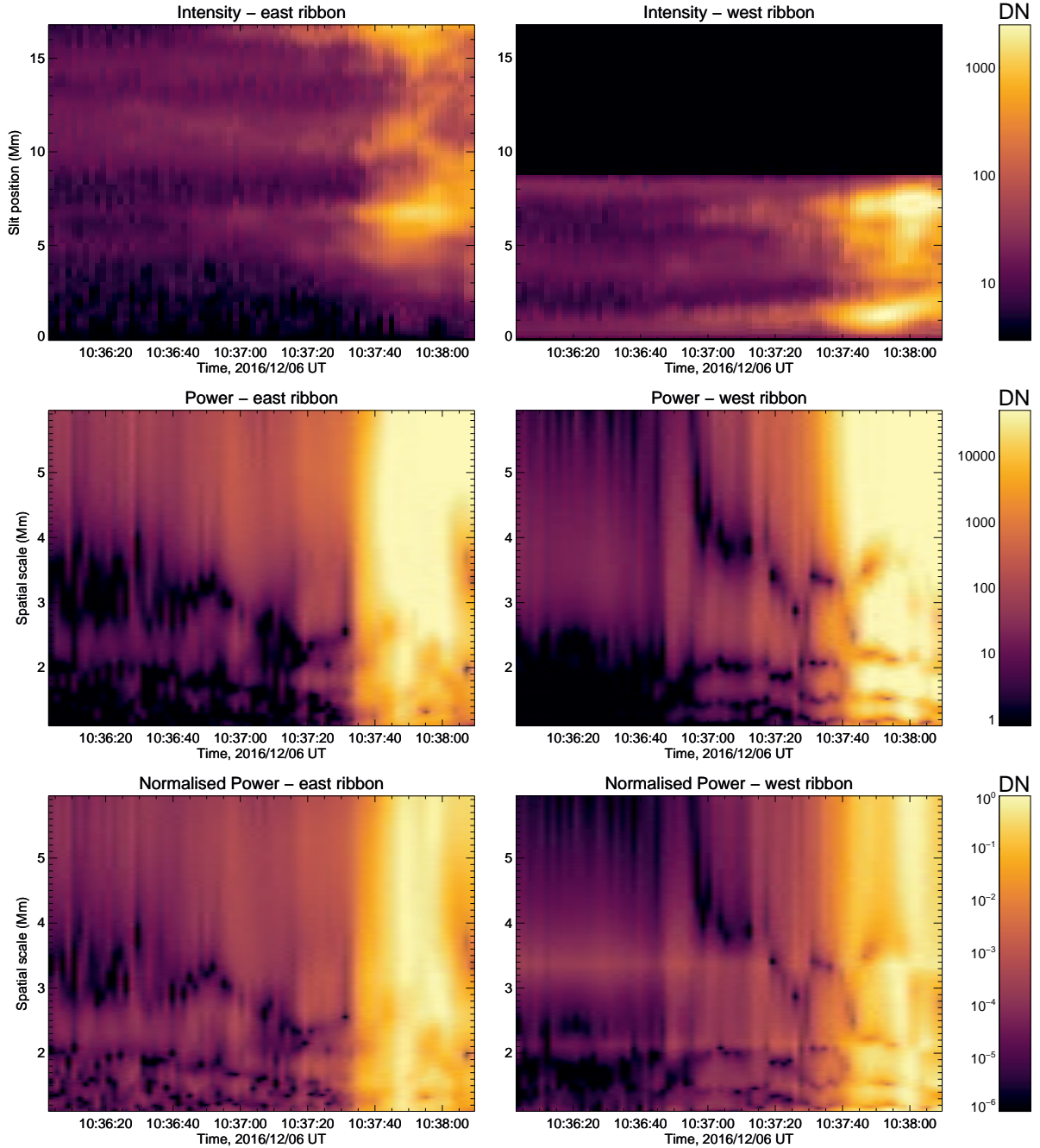}
  \caption{
Top: Ribbon intensity stack plots for both ribbons. 
Middle: Spatial scale power stack
plots. 
Bottom: Spatial scale power stack
plots, normalized at each spatial scale. All y-axes in units of Mm.
(0.33 arcsec). 
  }
  \label{fig:power}
\end{figure*}

We measure the intensity cross-section along the ribbons, by taking the mean intensity of a 3x3 pixel area around each pixel location in the centroid slit. Measuring an average around each pixel provides a more reliable estimate of intensity along the centroid (by not relying on a single pixel measurement), and reduces the impact of higher intensities in the slit-jaw image at the sit-and-stare slit location. Intensity measurements along this centroid are not uniformly spaced, dependent on whether adjacent pixels in the centroid lay horizontal/vertical or diagonally to one other. We interpolate along this centroid to create an evenly spaced array - necessary to calculate a Fast Fourier Transform (FFT). An example cross-section for each ribbon is shown in the bottom panel of Figure \ref{fig:tracking}. Here, we have converted position along the centroid into units of Mm. The east ribbon is over twice as long as the west, with centroid slit lengths of 17 and 8 Mm respectively. The intensity signal displayed here is then processed by detrending the data with a moving average, to avoid intensity jumps between the centroid slit ends dominating the power spectrum. The moving average sizes were selected manually, to remove large general trends in the data without inadvertently subtracting the smaller variations we aim to detect. Finally, we apply a Hanning Window to the data and introduce zero padding, to preserve the power and increase the spectral resolution of the Fast Fourier Transform (FFT). The FFT returns the power at spatial scales at intervals of L/2, L/3... L/n, where L is the length of the array in real units and n the number of data points. Zero padding therefore increases the spectral resolution of the FFT by calculating the power at smaller intervals (n is larger), as well as minimizing the effects of aliasing.

\subsection{2D Power Spectrum}

Intensity cross-sections along the ribbon centroid are measured and processed for every 1.7 s time step, producing the intensity stack plots in the top row of Figure \ref{fig:power}. For each of these cross-sections (vertical slices along the stack plot), we calculate the FFT of ribbon intensity, resulting in the power spectrum in the spatial domain for each time step. This provides us with the power of different spatial scales appearing along the flare ribbon. These spatial scales (in Mm) are inverse to the dispersion relation wavenumber, $k$, often used in the literature. We elect to work with units of spatial scale for this study, as they are easier to picture in this instance.
We combine each spatial FFT into a single stack plot, presented in the middle row of Figure \ref{fig:power} for each ribbon. Maximum power varies significantly with spatial scale, so to better see time evolution at each spatial scale we normalize the power spectrum in time, to produce the bottom panel of the same figure. Between 10:37 and 10:38 UT, we see a steep rise in power across all spatial scales, with growth rate and onset times varying across different scales. Before this time, we see consistent signals at certain spatial scales - around 2.5 Mm in the east ribbon (with a potential second, weaker peak higher around 5.5 Mm), and 2 and 3.5 Mm in the west ribbon. These signals result from the spacing between bright points in the pre-ribbon structure. This spacing can be verified by studying the intensity stack plots in the same figure, with the spacing between the bright horizontal features.

\section{Spatial scales and Exponential Growth}

\begin{figure}
  \centering
  \includegraphics[width=8.5cm]{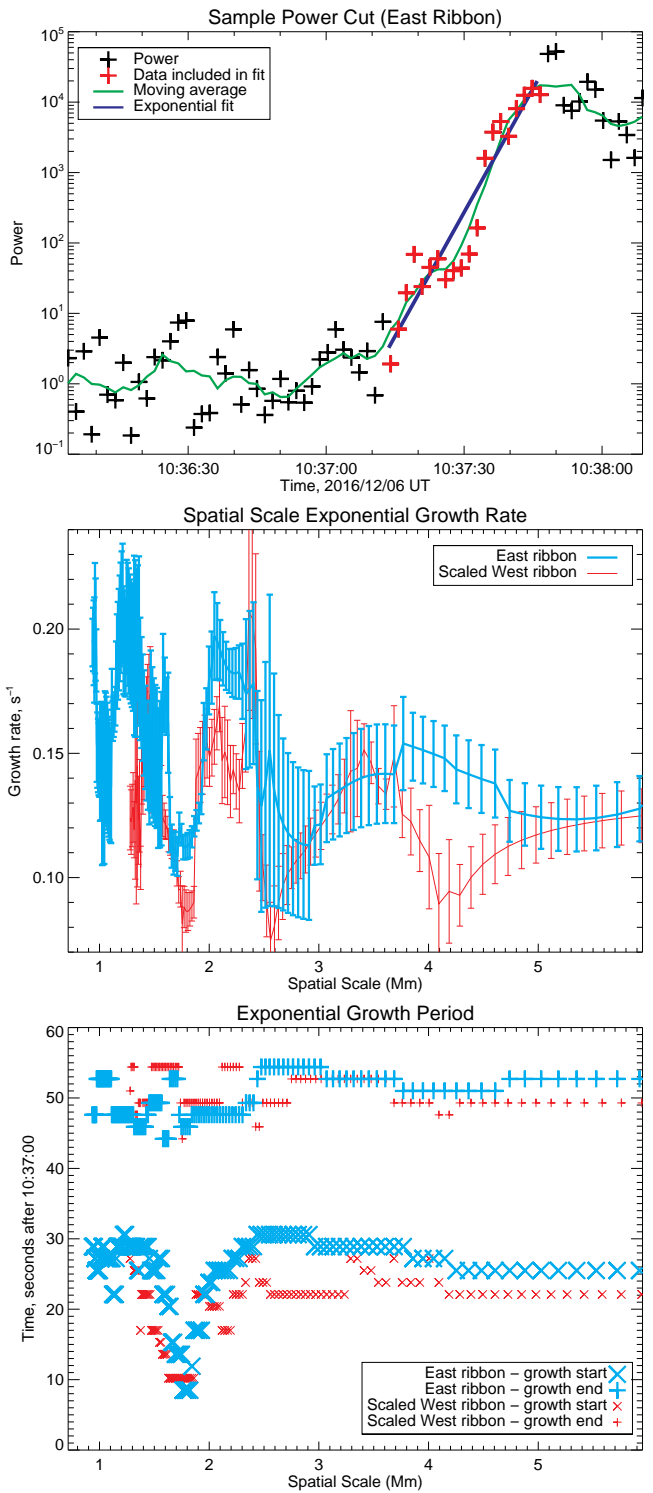}
  \caption{
Top: Cross-section through power plot (Fig 3) at spatial scale of 1.75 Mm, showing exponential growth. 
Middle: Exponential growth rate at
each spatial scale. West scales are transformed to east scales. 
Bottom:
Start/end times of the fitted exponential period, for the east and scaled west ribbon spatial scales.
}
  \label{fig:growth}
\end{figure}

\subsection{Flux Conservation - Spatial Scaling}

If the observed spatial scales in the flare ribbons result from processes propagating down from the reconnection site, we can expect the length scales to scale with the square root of the area of the flux tube connecting the two ribbons. (Picture a wave with one mode, perpendicular to the magnetic field direction. As the area of the flux tube changes, the wavelength scale approximately changes by the square root of the area). Considering the flux conservation shown in Figure \ref{fig:ribbons}, we can equivalently scale the length scales through the ratios of magnetic field strength (equation \ref{eqn:flux1}). 

As magnetic conditions vary along this tube of flux, we can apply this concept to compare the spatial scales seen at each ribbon, by scaling the west ribbon scales to the 'reference frame' of the east ribbon through the square root of the ribbon area ratios. The analysis below considers the original and scaled spatial scales observed in the east and west ribbon respectively.

\subsection{Growth with Spatial Scale}

In MHD, exponential growth at multiple spatial scales is a classical signal of plasma instability \citep{Priest1985}. We investigate the growth at a specific spatial scale by taking a horizontal cross-section through the power spectrum in Figure \ref{fig:power} - tracking the evolution of power with time. Figure \ref{fig:growth} (top panel) presents an example cross-section for a scale of 1.75 Mm in the east ribbon. The plot shows a noisy baseline power around $10^0$, which sharply increases by nearly 5 orders of magnitude between 10:37:10 and 10:37:50 UT. We determine the region of exponential growth in the data moving average, and fit an exponential curve to the corresponding data points. This provides us with an exponential growth rate and start/end times of the exponential phase.

By fitting an exponential curve for each spatial scale, we calculate the variation in the start time, duration and rate of the exponential growth of different spatial scales in each ribbon.
The east ribbon growth rates are shown as the cyan curve in the middle panel of Figure \ref{fig:growth}. The curve shows three peaks around 1.2, 2.3 and 3.8 Mm, before tailing off at high spatial scales. The error bars in this plot assume that the algorithm determining the start/end point of exponential growth was correct within one data point. With error bars considered, the plot is continuous with no major jumps in the data. The growth rate itself varies by a factor of $\sim 2-3$, between $\sim 0.09 - 0.24$ $s^{-1}$. The scaled west ribbon scales provide a good alignment with the east ribbon, and the location of the three peaks match reasonably well. 

We also plot the duration of exponential growth at each spatial scale for the east and west ribbon in Figure \ref{fig:growth} (bottom). Examining the east ribbon data, we see exponential growth start initially at a single spatial scale (the 1.75 Mm scale plotted previously), before beginning at all other scales up to 19 seconds later. This is suggestive of a process at a specific spatial scale causing the growth at progressively shorter and longer scales through a cascade and inverse cascade. The west ribbon produces noisier data (due to fewer data points along the ribbon) - but still shows exponential growth beginning first at the same spatial scale of 1.75 Mm. The end times of exponential growth are not captured as clearly by our algorithm. Detecting key spatial scales independently in both ribbons provides confidence that the processes causing the features are likely linked - that is, they both originate from instability processes at the reconnection site.

The east ribbon does not trace the strongest photospheric field of the active region, resulting in a fairly low mean LoS field strength of 107.5 Gauss. We do not expect the magnetic field of active region loops \citep{Brooks2021,Landi2021} to be significantly lower than this photospheric measurement. Considering again flux conservation, and that the length scales scale with the square root of the field strength ratios - the spatial scales of these processes at the reconnection site (in the active regions loops) will not be much larger than the scales observed in the flare ribbons (of order 1-10 Mm).

\section{Discussion}

\begin{figure*}
  \centering
  \includegraphics[width=16cm]{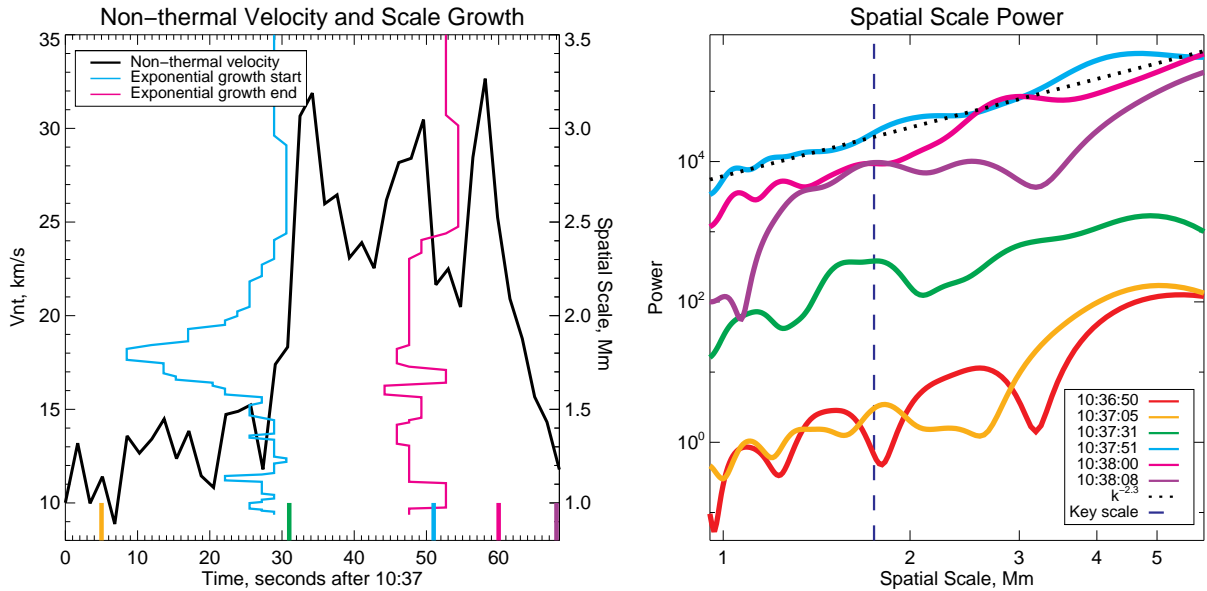}
  \caption{
 Left: Comparison between the non-thermal velocity timing, with the start an end of exponential growth at different spatial scales (in the east ribbon). Vertical ticks on the x axis mark the times shown in the right panel.
 Right: East ribbon spatial power spectra at key times, compared with a power-law distribution.
  }
  \label{fig:vnt}
\end{figure*}

Our analysis of intensity variations along the flare ribbons give an insight into the dynamics of the reconnecting current sheet. Our analysis has provided us with observational constraints, allowing us to compare the observed parameters with those predicted by reconnection instability theory. Our main constraints are -

\begin{itemize}
    \item Exponential growth is seen across all wavelengths, with preferred scale at 1.75 Mm.
    \item Exponential growth rate is in the order of $\approx 0.1-0.2$ $s^{-1}$.
    \item Spatial scale growth times suggest a cascade and inverse cascade.
\end{itemize}

In \citep{Tenerani2020}, the authors use 2.5D simulations to study the non-linear phase of tearing modes within a current sheet. The authors find growth beginning at a specific wave mode, before leading to growth at lower and higher modes (higher and lower spatial scales) through a simultaneous cascade and inverse cascade. The combination of cascades to higher and lower spatial scales simultaneously comes from the interplay of magnetic island collapse and coalescence. The exponential growth rates in \citet{Tenerani2020}, $\approx 0.45$ are also on the same order of magnitude as our observations. The specific properties of the tearing mode instability within this simulation are consistent with our observations. In particular, the cascade/inverse cascade, which we would not expect from other instability processes (such as Kelvin Helmholtz), suggests there is more at play than just turbulence. We therefore conclude that the flare ribbon spatial scale growth reflects the presence of the tearing mode instability in the current sheet.

With a higher spatial resolution, or larger flare with more data points along the ribbon - it would be possible to measure the power at further larger and smaller wavelengths. This would potentially allow for the measurement of a more complete dispersion relation, including the time variation ($\omega$) and second $k$ (variation across the ribbon) component, to allow further comparison with tearing mode theory.

\subsection{Flare timeline} 

Assuming that  exponential growth at our key spatial scale is the signature of tearing mode instability onset, we can compare this with the other observed parameters for this same event. \citet{Jeffrey2018} find that the rise in non-thermal velocity, believed to be a signature of turbulence at the flare ribbon slit location, preceded the intensity increase associated with plasma heating and flare onset. The authors note that there must be a driver for this long-lived turbulence signal. High-cadence IRIS observations of other small flares have found a similar pattern of enhanced non-thermal velocities preceding rises intensity - consistent with signatures of turbulent reconnection in the current sheet \citep{Chitta2020}. Comparing the timing of the non-thermal velocities with the growth of spatial scales in Figure \ref{fig:vnt} (left),  we find growth at our key scale of 1.75 Mm begins some 15 seconds before the growth in non-thermal velocities. Similarly, the end of exponential growth across all spatial scale occurs a similar time frame before the drop of non-thermal velocity. This suggests that for this event, the tearing mode instability is the driver of plasma turbulence in the region, as a result of the cascade and inverse cascade to progressively smaller and larger spatial scales. 

Further evidence to support the breakdown to turbulence via a(n) (inverse) cascade can be found by examining the power spectra for specific time frames (taking a vertical cross-section through the middle row of Figure \ref{fig:power}). Cross-sections at important moments are shown in Figure \ref{fig:vnt} (right), and are as follows:

\begin{itemize}
    \item 10:36:50 - Large oscillations in the low-amplitude power spectra, equivalent to the harmonics of the bright points in the pre-flare ribbon region.
    \item 10:37:05 - Key scale of 1.75 Mm appears for the first time.
    \item 10:37:13 - Exponential growth begins at the key scale (shown in left panel of Figure \ref{fig:vnt}).
    \item 10:37:31 - Exponential growth is well underway across all spatial scales, lead by the key 1.75 Mm scale. The non-thermal velocity begins to rise at this time.
    \item 10:37:51 - Power spectra have reached their peak, with all scales catching up to the key scale to produce a power law of $\approx k^{-2.3}$.
    \item 10:38:00 onwards - Power law starts to break down, with a falling amplitude and local maxima/minima forming. Non-thermal velocity is also returning to background levels.
\end{itemize}

At the peak of the power spectra, the spatial scales form a power law of approximately 2.3. A power law in this domain is a key prediction in plasma turbulence, routinely seen in in-situ measurements. In the classic Kolmogorov-like distribution of turbulence \citep{Kolmogorov1941}, the power law is predicted to evolve to and end-state of 5/3 - slightly lower than our observed value of $\approx 2.3$.  There are two likely explanations for this disparity. Firstly, it is possible that the tearing mode process did not persist for long enough to drive the power law to the predicted Kolmogorov  end-state of 5.3. Or, an alternative turbulence model is a better fit for these conditions. \citet{Tenerani2020}, for example, predict a power-law index of 2.2, when introducing a model of intermittency \citep{Frisch1978} to the collapsing magnetic islands and beginning from a Kolmogorov state. \citet{Dong2018} also find an index of 2.2, in their simulation of a plasmoid-mediated regime.

By comparing the power spectrum and non-thermal velocity evolution with the timing of exponential growth onset, we have provided evidence for the tearing-mode instability manifesting itself before plasma turbulence - similar to that expected by simulation work, e.g. \citet{Dong2018,Tenerani2020}. As shown in \citet{Jeffrey2018}, the turbulence itself precedes the ribbon intensity enhancements and flare onset. We thus have extended our knowledge of the pre-flare timeline for this single event, and provided insight into the nature of current-sheet collapse and onset of fast-reconnection via MHD instabilities in a solar flare.

\section{Conclusion} 
A major open question in solar flare physics is the sequence of events that lead to the onset of the fast reconnection that is widely acknowledged to cause the impulsive energy release. Central to this question is the formation and disruption of the current sheet in which reconnection occurs. Current sheet formation can occur through a variety of processes that include large scale MHD instabilities that are a key element of eruptive flare models \citep[e.g.][and references therein]{Green2018}, but also as the result of MHD turbulence. The subsequent evolution of that turbulence can then lead to its disruption through the onset of the tearing mode instability, which causes the energy spectrum to steepen \citep{Dong2018}. The issue of the interplay and feedback between reconnection, turbulence and current sheet disruption is therefore complex, but critical to understanding flare onset.

For the event studied here we find evidence to support the development of the tearing mode instability prior to the first detectable signatures of turbulence, followed by the subsequent evolution of the energy spectrum to a power-law dominated regime with slope $\approx 2.3$. While we cannot completely rule out the existence of turbulence prior to the tearing mode onset that we are just unable to detect with the available data, our analysis strongly suggests current sheet formation by another mechanism, e.g. at the interface between emerging and pre-existing magnetic fields. The subsequent evolution of the instability and turbulent energy spectrum is consistent with previous work by \citet{Dong2018} and \citet{Tenerani2020} that supports the presence of the combined effects of dynamic alignment of the magnetic islands (or plasmoids), intermittency and recursive reconnection. We note that recent work by \citet{French2019,French2020} provides an important new tool for quantifying the internal sub-structure of current sheets that can be used to further probe alignment and intermittency.

In conclusion, we find that for case study presented here the flare onset is precipitated by current sheet formation, leading to the onset of the tearing mode instability and subsequent development of a turbulent cascade.

\acknowledgments
IRIS is a NASA small explorer mission developed and operated by LMSAL with mission operations executed at NASA Ames Research Center and major contributions to downlink communications funded by ESA and the Norwegian Space Centre.
R.J.F. thanks the STFC for support via funding from the PhD Studentship, ST/S50578X/1. 
S.A.M was supported by STFC Consolidated Grant (CG) ST/S000240/1 and Hinode Operations (UKSA) ST/S006532/1.
I.J.R was supported in part by ST/V006320/1. 
AWS was supported by CG ST/S000240/1, and NERC NE/P017150/1 and NE/V002724/1.
The authors thank Gherardo Valori, Lidia van Driel-Gesztelyi, Clare Watt and Philip Judge for their helpful discussions.

\bibliographystyle{aasjournal}
\bibliography{bibliography}

\end{document}